\documentclass[sigconf]{acmart}

\AtBeginDocument{%
  }

\setcopyright{acmlicensed}
\copyrightyear{2025}
\acmYear{2025}
\acmDOI{XXXXXXX.XXXXXXX}

\acmConference[]{}{}{}

\usepackage{subfig}
\usepackage{booktabs}
\usepackage{eso-pic}
\usepackage{lipsum}
\usepackage{listings}
\usepackage{pifont}
\usepackage{wasysym}
\usepackage[framemethod=TikZ]{mdframed}
\usetikzlibrary{shadows}
\usepackage{subfig}
\usepackage{multirow}
\usepackage{makecell}
\usepackage{threeparttable}
\usepackage{balance}
\usepackage{marvosym}
\usepackage{setspace}
\usepackage[htt]{hyphenat}
\usepackage{enumitem}
\usepackage{color,soul}
\usepackage{mysettings}

\renewcommand\footnotetextcopyrightpermission[1]{}
\begin{document}

\AddToShipoutPictureBG*{%
  \AtPageUpperLeft{%
    \hspace{\paperwidth}%
    \raisebox{-\baselineskip}{%
      \makebox[-12pt][r]{\textbf{SAND2025-03552C}}
}}}%
\title[Knowledge Transfer from LLMs to Provenance Analysis: A Semantic-Augmented APT Detection]{Knowledge Transfer from LLMs to Provenance Analysis:\\A Semantic-Augmented Method for APT Detection}


\author{Fei Zuo}
\email{fzuo@uco.edu}
\affiliation{%
  \institution{University of Central Oklahoma}
  \city{Edmond}
  \state{Oklahoma}
  \country{USA}
}

\author{Junghwan Rhee}
\email{jrhee2@uco.edu}
\affiliation{%
  \institution{University of Central Oklahoma}
  \city{Edmond}
  \state{Oklahoma}
  \country{USA}
  }

\author{Yung Ryn Choe}
\email{yrchoe@sandia.gov}
\affiliation{%
  \institution{Sandia National Laboratories}
  \city{Livermore}
  \state{California}
  \country{USA}
}

\renewcommand{\shortauthors}{}

\begin{abstract}

Advanced Persistent Threats (APTs) have caused significant losses across a wide range of sectors, including the theft of sensitive data and harm to system integrity. As attack techniques grow increasingly sophisticated and stealthy, the arms race between cyber defenders and attackers continues to intensify. The revolutionary impact of Large Language Models (LLMs) has opened up numerous opportunities in various fields, including cybersecurity. An intriguing question arises: can the extensive knowledge embedded in LLMs be harnessed for provenance analysis and play a positive role in identifying previously unknown malicious events? To seek a deeper understanding of this issue, we propose a new strategy for taking advantage of LLMs in provenance-based threat detection. In our design, the state-of-the-art LLM offers additional details in provenance data interpretation, leveraging their knowledge of system calls, software identity, and high-level understanding of application execution context. The advanced contextualized embedding capability is further utilized to capture the rich semantics of event descriptions. We comprehensively examine the quality of the resulting embeddings, and it turns out that they offer promising avenues. Subsequently, machine learning models built upon these embeddings demonstrated outstanding performance on real-world data. In our evaluation, supervised threat detection achieves a precision of 99.0\%, and semi-supervised anomaly detection attains a precision of 96.9\%.


\end{abstract}

\ccsdesc[500]{Security and privacy~Intrusion detection systems}
\ccsdesc[300]{Computing methodologies~Machine learning}

\keywords{Intrusion Detection, APT Detection, Provenance Analysis, GPT, Large Language Models}

\maketitle

\section{Introduction}

In recent years, advanced persistent threats (APTs) targeting key sectors such as government, finance, and business have been on the rise. These cyberattacks employ increasingly complex techniques, are prolonged in duration, and are difficult to detect, resulting in significant economic losses. As early as 2017, APT actors breached the network of the credit reporting agency Equifax, ultimately resulting in losses exceeding 425 million US dollars~\cite{equifax_case}. According to CrowdStrike's report~\cite{threat_report}, ``\textit{cloud environment intrusions increased by 75\% from 2022 to 2023}''. Meanwhile, the number of victims named on dedicated leak sites increased by 76\%, indicating that data-theft extortion remains rampant. 

Among the emerging technologies for robust APT detection, system provenance analysis is being considered as a promising mechanism, thus attracting widespread attention. As cyber threats become more complex and frequent, traditional approaches to threat detection and response are proving inadequate. Therefore, we also notice that leveraging the progress in AI to assist and automate system provenance analysis has become more of a need than an option. 
Industry statistics show that in practical incident response applications, fully deployed AI-driven systems
``\textit{were able to identify and contain a breach 28 days faster than those that didn’t, saving USD 3.05 million in costs}'' for the organizations~\cite{ibm2023}. The recent survey~\cite{ai_security} indicates that ``\textit{70\% of cybersecurity professionals believe AI is highly effective in detecting previously undetectable threats}''. Additionally, 73\% of cybersecurity teams want to shift their focus to an AI-powered preventive strategy. 

\begin{lstlisting}[
  label={list_app01},
  caption={The explanation of a system event (expressed using a triplet ``\texttt{vim read /etc/localtime}'') provided by GPT-4.},
  captionpos=b,
  frame=shadowbox, 
  basicstyle          =   \ttfamily\footnotesize,
  keywordstyle        =   \bfseries, 
  rulesepcolor=\color{red!20!green!20!blue!20},
  keywordstyle=\bfseries, 
  showstringspaces=false,
  breakatwhitespace=false,         
  breaklines=true,       
  keepspaces=true,
  xleftmargin=8pt,
  xrightmargin=10pt,
  morekeywords={vim, read, etc, localtime, // }
]
 The text editor vim performed a read operation to 
 access the system file /etc/localtime, which is 
 typically used to retrieve timezone configuration 
 information.
\end{lstlisting}
\settopmatter{printfolios=true}

Existing AI-powered threat detection techniques have either approached provenance analysis from a graph perspective or from a natural language perspective~\cite{cheng2024kairos, wang2020you}. In this work, we follow the latter approach, considering the system entities and the interaction between them as various components of a sentence. 
But unlike previous work, we will take advantage of the recent advancements in large language models (LLMs) to enhance provenance analysis. This approach offers several obvious benefits. First of all, \textbf{the extensive knowledge possessed by LLMs can be leveraged to augment the semantics of system event descriptions}. For example, given a simple system event described by the triplet ``\texttt{vim read /etc/localtime}'', the descriptive text generated by GPT-4 through being guided by appropriate prompt engineering is shown in Listing~\ref{list_app01}. Compared to the original input triplet, GPT-4 accurately supplemented the information that ``\texttt{vim} is a text editor''. Moreover, based on the characteristics of the second system entity, it inferred that the purpose of this file reading behavior was to obtain timezone configuration information. Previous research~\cite{ming2017binsim} has shown that understanding such information is highly useful when analyzing malicious behavior by attackers. This is because some adversaries deliberately exclude certain areas from their attacks. For instance, a family of immensely dangerous and destructive crypto-ransomware are used to first checking the infected machine's UI language before launching the attack. If the language is Russian, Ukrainian, or another language from former Soviet Union countries, the attack will not be triggered. However, without the additional information offered by LLMs, similar tasks would typically require significant human effort and expertise. More examples will be provided later in Section 3.3. 


It is worth noting that the descriptive texts used to interpret system events are unstructured data. Therefore, to facilitate subsequent learning by artificial intelligence models, we also need to use appropriate embedding models to learn numerical representations of data. This highlights another benefit of introducing LLMs: \textbf{their powerful embedding models can accurately capture and retain more semantic information}. In the natural language processing domain, different embedding techniques have been proposed
for words~\cite{mikolov2013distributed}, sentences~\cite{pagliardini2018}, and documents~\cite{le2014distributed}. The prior work~\cite{wang2020you} used \texttt{Doc2Vec}~\cite{le2014distributed} to generate embeddings for a series of selected system events. However, when using traditional models like \texttt{Word2Vec}~\cite{mikolov2013distributed} or \texttt{Doc2Vec}~\cite{le2014distributed}, the out-of-vocabulary (OOV) issue will be an unavoidable challenge. For provenance analysis in the wild, it is unrealistic to exhaustively collect words such as file names or executable paths. A usual approach is to ignore words that have never appeared in the vocabulary when generating numerical representations of sentences. However, this leads to information loss. In our proposed method, we offer two targeted solutions. First, for system entities with indicators that have special significance, we make full use of the extensive knowledge of LLMs to interpret them. Second, for entities like system-generated temporary files or hash values, we have designed a heuristic approach to preprocess them. We will show more details in Section 3.

\begin{figure}[!t]
\centerline{\includegraphics[width=0.45\textwidth]{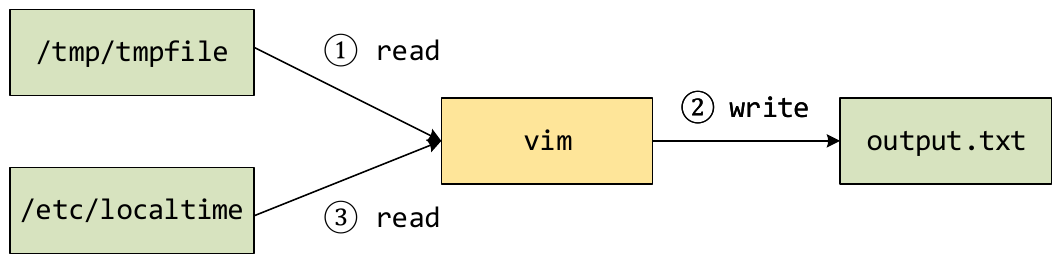}}
\caption{A simplified provenance graph example.}
\label{prov_eg}
\end{figure}

Though we have witnessed a number of meaningful attempts to integrate LLMs into various cybersecurity applications~\cite{yan2023prompt, qu2024context, zhang2024prompt}, the extent to which LLMs can assist in provenance analysis remains an open question. In this work, we aim to gain a deeper understanding of the potential of LLMs in enhancing provenance analysis. To this end, we developed a proof-of-concept prototype which fully takes advantage of the extensive knowledge of LLMs. Our approach leverages LLMs to augment the semantics of system events' descriptions, and further generate high-quality embeddings. In the later threat detection phase, we comprehensively consider supervised and semi-supervised anomaly detection models. Our evaluation was conducted based on a publicly accessible dataset, with samples derived from real-world attacks. The experiment results show that our proposed technique is able to distinguish malicious events from benign ones, maintaining a high level of accuracy even the attack was previously unseen.

The main contributions of our work are as follows:
\begin{itemize}

\item We have explored a systematic strategy for applying the extensive knowledge of LLMs in APT detection through provenance analysis. Our innovative approach leverages LLMs to augment system event descriptions and further utilizes the assistance of LLMs for high-quality representation learning.

\item We conducted comprehensive studies on the performance of the advanced contextualized embedding approach 
in learning the semantics of system event descriptive texts. Subsequently, both the supervised and unsupervised detection methods built based on this achieved high precision.

\item Our study utilizes a dataset collected from real-world cyber-attacks. Based on this, we showcased the capability of LLMs in improving provenance analysis through extensive experiments. The research resources are available upon request to facilitate future related studies.



\end{itemize}

The remainder of this paper is organized as follows: First, to help readers better understand this article, we briefly introduce the necessary preliminary knowledge in Section~\ref{sec:bck}. Next, Section~\ref{sec:method} describes our system architecture and design in detail. Then, we conduct evaluations and show the results in Section~\ref{sec:eval}. The related work is then reviewed in Section~\ref{sec:related}. The limitations of this work and potential future directions are discussed in Section~\ref{sec:limit}. Finally, we draw conclusions in Section~\ref{sec:conclude}.

\section{Background}\label{sec:bck}

In this section, we briefly introduce the necessary background on system-level provenance analysis and large language models.

\subsection{Provenance Data}

Provenance data records complex dependencies across various system events, thus reflecting the interactions between entities within a historical context. System-level provenance is usually presented in the form of graphs, where each node represents a system entity (e.g., a process or a file); edges are system call and timestamp labels related to the nodes. A system call is the programmatic way in which a user app can request a service from an OS kernel.

\begin{figure*}[!th]
\centerline{\includegraphics[width=0.9\textwidth]{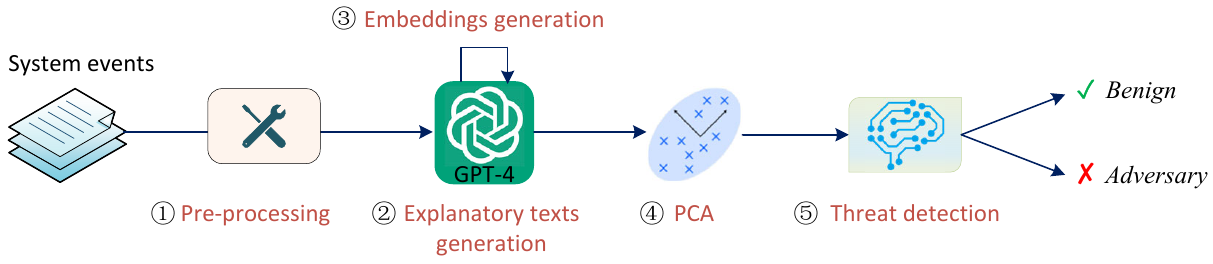}}
\caption{System overview.}
\label{sys_overview}
\end{figure*}

Figure~\ref{prov_eg} shows a provenance graph example, which includes four system entities represented by the four nodes in the graph. An entity refers to a component within a system responsible for producing, modifying, or processing information and resources. The operation exerted by one system entity on another is described using a system call, for instance, \texttt{read} and \texttt{write} in Figure~\ref{prov_eg}. It should be noted that system provenance graphs in reality are very complex and extensive; herein, we only use a simplified example for demonstration purposes. Provenance data is often represented as a graph because graphs can intuitively depict dependencies between events or their chronological order. However, considering that hackers launch multi-stage attacks, extracting attack-relevant events from a vast number of system events over a long time span and linking them is non-trivial. As a result, the generated provenance graph often contains significant noise, posing challenges for attack detection. Therefore, it should be noted that \textbf{the proposed approach focuses on system events} in provenance data.

The two interrelated system entities, along with the operation between them, collectively constitute an event. For example, we can observe three system events in Figure~\ref{prov_eg}. In temporal order, they are (1) the text editor \texttt{vim} reads a temporary file; (2) the process \texttt{vim} saves its content to a file named \texttt{output.txt}; and (3) \texttt{vim} checks the system's time zone by reading the \texttt{/etc/localtime} file. A system event is a basic unit in provenance data for tracking and recording system-level behaviors. Hence, system events extracted from a provenance data are regarded as informative features for describing a cyber incident in intrusion detection. 

\subsection{Large Language Models}

In recent years, the emergence of large language models (LLMs) is undoubtedly one of the most impactful breakthroughs of AI. A large language model, as the name suggests, is a language model trained on massive amounts of text data and has billions of parameters. As a widely recognized instance of an LLM, the Generative Pre-trained Transformer (GPT) has attracted widespread attention worldwide since its emergence in 2018~\cite{radford2018improving}. As an implementation of GPT, OpenAI launched ChatGPT\footnote{\url{https://chatgpt.com/}} to the public in 2022, which demonstrates a powerful ability to generate human-like responses in conversational interactions. Microsoft's Copilot\footnote{\url{https://copilot.microsoft.com/}} is another representative commercial product of generative AI, which adopts GPT as the core model. Another main competitor in the same field, Gemini\footnote{\url{https://gemini.google.com/}} was launched by Google in December 2023. 


In this work, we chose GPT as our primary LLM because OpenAI has also released a corresponding API for external users to access the advanced LLMs developed by OpenAI. Applying the API offered by OpenAI in applications has three main benefits. First, developers are allowed to interact with a vast knowledge base provided by OpenAI's state-of-the-art AI models in a programmatical manner. Second, it provides a straightforward way to leverage the cutting-edge capabilities in NLP, thus enabling tasks like text generation, interpretation, and summarization. Lastly, other functionalities provided by OpenAI, such as fine-tuning, facilitate users to optimize their application's performance on customized tasks.

\subsection{Unstructured Data Embeddings}

Provenance data is recorded in an unstructured form, such as file names or executable paths, which are textual in nature. To facilitate subsequent machine learning tasks, embedding methods are required to learn the numerical representations for the unstructured data. Multiple embedding techniques have been proposed in the field of natural language processing. In some previous security-related applications~\cite{zuo2024vulnerability, zuo2023commit}, the practice of embedding text has played a crucial role. Currently, mainstream embedding methods can be categorized into two main types: static embeddings and contextualized embeddings. When using static embeddings, the same word always has the same representation, even when it appears in different contexts. For example, the word ``mouse'' can refer to either a computer hardware or an animal, but regardless of the context, it only has one vector representation. In contrast, contextualized embeddings generate dynamic, context-dependent vector representations for text, meaning that the representation changes depending on the surrounding context.

The strength of analogical reasoning ability is an important indicator of whether text embeddings are meaningful. An analogical question typically consists of two pairs of texts, such as (``man'', ``king'') and (``woman'', ``queen''). To assess how related the two pairs are, the analogy ``man is to king as woman is to queen'' is formed, and its validity is tested. In the feature space, the analogical question can be represented as $ |E_1 - E_2| \approx |E_3 - E_4| $, where $E_1$, $E_2$, $E_3$, and $E_4$ are the embedding vectors, and $|\cdot|$ represents the norm. This means that the two pairs of texts being examined can form an approximately parallelogram-like relationship in the feature space. Applying this concept to our problem, we can find similar analogies between one pair of system event descriptions and others. If these analogies align with our prior knowledge expected in provenance analysis, we can confirm that the embeddings generated by LLMs are semantically meaningful.

\section{Methodology}\label{sec:method}

In this study, we employ OpenAI API to access and integrate the capabilities of GPT-4o into our applications. GPT-4o is a state-of-the-art multimodal LLM created by OpenAI, which has been trained on data up to October 2023, ensuring its responses are informed by the most recent and relevant information available. 
Through the API, we can automatically interact with GPT-4o to perform various tasks, such as generating text in batches. 

Figure~\ref{sys_overview} illustrates the overall pipeline of our system.
Given the provenance system data, we first pre-process the raw data according to the specific characteristics of different types of system events. Second, using the LLM service, explanatory texts with augmented semantics for system events are generated with proper prompt engineering. Third, the text embeddings are further produced by taking advantage of the facilities of the LLM. Next, we employ some manifold learning method such as Kernel PCA to adjust the dimensionality of the original embeddings. Lastly, the threat is detected by a well-trained machine-learning model.

\subsection{Pre-processing}

\textbf{Input data:}
We use a system call event whose several properties are extracted such as the system call name, user name, user shell, process name for all events. The file descriptor name is included for file events.
For process creation events such as \texttt{clone} and \texttt{execve}, system call arguments are included because they include specific information about what program is executed. This information is essential for interpreter programs such as Python and Bash because the main program names do not show the script names. For network events, network type (IPv4), client IP, server IP, and server port are used. These fields are provided as a JSON object to the LLM service.


\textbf{Path normalization:}
Using the raw data sometimes could create unnecessary noise to the LLM as seen in the randomly generated data or timestamps. Therefore, we applied several data normalization methods to avoid such noise. First, temporary files tend to have less meaningful names due to their nature. We utilized several well-known file extensions for the temporary files such as ``.tmp'' and ``.temp'' to generalize their names. 
Second, the files under certain directories such as the \texttt{proc} file system can be noisy because they reflect the internal states of the operating system kernel and their file paths may include the process IDs (PIDs) which are dynamically determined. Therefore, such file names are generalized by excluding the PIDs.
Lastly, if file names are like hashes based on known patterns (e.g., MD5 and SHA), their names are normalized using a uniform pattern ``hash value''.

\subsection{Explanatory texts generation}

The request messages are structured to
include two types of content: one for the ``system'' role and another for the ``user''
role. The ``system'' role content outlines general instructions, detailing the input
required from the user and the expected output from GPT-4o. Meanwhile, the
``user'' role content provides the details of a system call as a query.

\textbf{System role:} For the ``system'' role, we used a prompt \textit{``here is a json object where the keys are sysdig event fields and the values are the data for the corresponding keys. Please generate only the final summary of the meaning of this event and the event type as a paragraph''}. Our implementation of system call events utilizes the \texttt{Sysdig}~\cite{sysdig2024} system which happens to be recognized by the LLM service that we use. This is one example of system call formats and other formats can be utilized as well by instructing the LLM on how to understand them accordingly. This prompt describes that we are providing our data in the JSON format and we would like to have a summary of the meaning of this event. Since the LLM may use verbose expressions, we instruct to produce only the final summary as a paragraph.

\textbf{User role:} The user role uses the details of a system call's properties as the key-value pairs in the JSON format.

\begin{lstlisting}[
  label={list_json},
  language=python,
  caption={A JSON example for the user role.},
  captionpos=b,
  frame=shadowbox, 
  basicstyle          =   \ttfamily\footnotesize,
  keywordstyle        =   \bfseries, 
  stringstyle=\color{numbercolor},
  rulesepcolor=\color{red!20!green!20!blue!20},
  keywordstyle=\bfseries, 
  showstringspaces=false,
  breakatwhitespace=false,         
  breaklines=true,       
  keepspaces=true,
  xleftmargin=8pt,
  xrightmargin=10pt
]
{
    'proc_name': 'crond', 
    'type': 'execve', 
    'fd_filename': '<NA>', 
    'user_name': 'root', 
    'user_shell': '/bin/bash', 
    'evt_args': 'sh'
}
\end{lstlisting}

Listing \ref{list_json} shows an example of a system event which is made by the \texttt{crond} process executing the \texttt{sh} program listed in the \texttt{evt\_args} system call parameter list field using the \texttt{execve} system call.
Since this is a process event, the \texttt{fd\_filename} field is not used as shown by its value \texttt{<NA>}. This system call is executed by the \texttt{root} user whose shell is \texttt{/bin/bash}.

Given these input data, the OpenAI \texttt{gpt-4o} LLM service responded with the following answer:
``\textit{This event is an \texttt{execve} type, indicating the execution of a new program. The process involved is \texttt{crond}, which is a daemon used to execute scheduled commands. The event was initiated by the \texttt{root} user, utilizing the \texttt{/bin/bash} shell. The specific command executed is \texttt{sh}, which suggests the initiation of a shell session or script execution. This activity is typical for system maintenance or scheduled tasks managed by the \texttt{cron} service.}''

This text explanation of a system event is used as an input for the text embedding explained next.

\subsection{Notable Details in LLM Explanations}

During our exploration of LLM explanations of system call events, we identified that there are multiple notable details that are beneficial for understanding the events, especially for cybersecurity purposes. We list several concrete examples as demonstrations.

\begin{enumerate}[]

\item \textbf{Knowledge of System Calls:} First, in LLM's explanation, a simple system call name is expanded to a full sentence of what it indicates by explaining what is the system call for.
({\setlength{\fboxsep}{0pt}\colorbox{babyblue}{blue}} annotation)

\item \textbf{Knowledge of Software Identity:} Second, For commonly used software, LLM is aware of what software it is. It helps the understanding of its existence or expected behavior if the user lacks knowledge.
({\setlength{\fboxsep}{0pt}\colorbox{myviolet}{violet}} annotation)

\item \textbf{High-Level Knowledge of Application Execution Context:}
Based on the prior two types of knowledge, LLM may suggest what is the current behavior doing in a higher-level description that is easier to understand to human. 
({\setlength{\fboxsep}{0pt}\colorbox{bananamania}{yellow}} annotation)

\item \textbf{Comment on Possible Suspiciousness:}
Lastly, LLM sometimes comments on possible usages of the current behavior and potential suspicious cases that is worth for the attention for possible security threats. This is a useful extra knowledge extracted from the LLM. ({\setlength{\fboxsep}{0pt}\colorbox{pink}{pink}} annotation)

\end{enumerate}

In the following examples, we explain the LLM's knowledge with the four types of details aforementioned.

Example 1 shows the explanatory text generated by the LLM for the system event ``\texttt{dhclient recvfrom root /bin/bash <NA> IP2 IP1 67}''.  It is worth noting that although we adopt the JSON format to describe a system event when interacting with LLMs, herein, for the sake of convenience and brevity, we use a textual format to represent a system event. Unless otherwise specified, all subsequent examples in this paper will follow this approach.

\mdfsetup{shadow=true,
shadowsize=5pt,
backgroundcolor=gray!10,
frametitlebackgroundcolor=yellow!20,
frametitlerule=true,
roundcorner=6pt,
skipabove=6pt,
skipbelow=6pt, 
rightmargin=15pt
}
\begin{mdframed}[frametitle={Example 1}]
This event is a network activity event involving the process \texttt{dhclient},\codeviolet{which is a DHCP client typically used to obtain IP addresses and network configuration from a DHCP server.} The event type is \texttt{recvfrom},\codeblue{indicating that the process received data from a network socket.} The data was received over an IP v4 connection from the source IP address \texttt{IP1}, which is likely the DHCP server, using the source port 67,
  \codeback{the standard port for DHCP server communications.}
  The destination IP address is \texttt{IP2}, which appears to be a broadcast address. The process is running under the user \texttt{root} with the shell \texttt{/bin/bash.}\codeback{This event suggests that the DHCP client is receiving network configuration information from the DHCP server.}
\end{mdframed}

\textbf{Manual Evaluation:} The first example is about a \texttt{dhclient} process performing a \texttt{recvfrom} system call event. LLM first explains the identity of this software correctly as a DHCP client annotated with violet color. It adds the explanation of what this system call indicates (blue annotation). Then the LLM explains a high-level application context of this event as the DHCP client receiving network configuration information from the DHCP server (yellow annotation).

Given the system event ``\texttt{screen openat ubuntu /bin/bash LC\_TIME}", Example 2 shows the corresponding  explanatory text generated by the LLM.

\begin{mdframed}[frametitle={Example 2}]
This event is an \texttt{openat} system call\codeblue{where the process named \texttt{screen} is attempting to open a file or directory named \texttt{LC\_TIME}}. The action is performed by a user named \texttt{ubuntu} who is operating within a Bash shell environment.\codeback{This type of event typically indicates a process accessing or modifying locale settings related to time, possibly for configuring or retrieving time-related information.}
\end{mdframed}

\textbf{Manual Evaluation:} 
The second example is about the \texttt{openat} system call event made by the \texttt{screen} software. LLM explains that this system call is regarding opening a file or directory correctly (blue annotation). Then it adds a high-level description that it is typically about accessing or modifying locale settings related to time, which is inferred by the accessed file, \texttt{LC\_TIME}. (yellow annotation)

The last system event used as a sample is ``\texttt{echo mprotect www-data /usr/sbin/nologin <NA>}''. The corresponding explanatory text generated by the LLM is shown in Example 3.

\begin{mdframed}[frametitle={Example 3}]
This event is of type \texttt{mprotect} and involves the process \texttt{echo} being executed by the user \texttt{www-data,} who has a restricted shell set to \texttt{/usr/sbin/nologin.}\codeblue{The \texttt{mprotect} system call is typically used to change the memory protection of a region of memory, which can be indicative of a process attempting to modify its memory permissions.}The absence of a specific file descriptor suggests that this operation is not directly associated with a file. This event could be part of a legitimate operation or\codepink{potentially suspicious activity, depending on the context, as \texttt{www-data} is a common user for web services, and the restricted shell indicates limited direct user interaction.}
\end{mdframed}

\textbf{Manual Evaluation:} 
The last example is about the \texttt{echo} process executing the \texttt{mprotect} system call. LLM first explains the meaning of this system call changing the memory protection of a region of memory.
We found a more useful description is about LLM's comment on this event's usage. LLM mentions this event could be part of a legitimate operation or potentially suspicious given a web-service user with a restricted shell.
Indeed, this \texttt{echo} process was caused by a security exploit from a malicious behavior dataset and LLM correctly identified its potential risk.


\subsection{Embeddings and Dimension Adjustment}

As we mentioned earlier, \texttt{ProvDetector}~\cite{wang2020you} uses \texttt{Doc2Vec}~\cite{le2014distributed} to embed the descriptive text of system events. \texttt{Doc2Vec}~\cite{le2014distributed} is a static embedding method that learns paragraph embeddings via the distributed memory and distributed bag of words models. In contrast, OpenAI has launched two powerful embedding models since January 2024, namely \texttt{text-embedding-3-small} and \texttt{text-embedding-3-large}~\cite{emb_openai}. In particular, they use the state-of-the-art contextualized embeddings method, which is considered as an augmentation to static embeddings. 

Static embeddings assign a single and fixed representation to each word, regardless of its context. However, contextualized embeddings generate more accurate and flexible representations by considering the context in which words appear, thus providing fine-grained semantics understanding and more nuanced representations. In the NLP domain, contextualized embeddings have led to substantial performance gains on a variety of tasks compared to static embeddings. The embedding models of OpenAI enable us to generate numerical sequences that accurately capture the semantics of natural language, laying a solid foundation for subsequent machine learning tasks.

OpenAI's embedding models can output feature vectors with dimensions as high as 1,536 or 3,072. Generally, higher-dimensional feature vectors can represent more detailed and complex information. However, high-dimensional feature vectors also pose challenges for subsequent machine learning model training. A common approach is to reduce the dimensionality of the feature vectors before the next stage of machine learning model training. Dimensionality reduction inevitably leads to some loss of information, so a trade-off must be made between time cost and information loss. In this work, we use the Kernel Principal Component Analysis (PCA) with an RBF (Radial Basis Function) kernel to compress the high-dimensional feature vectors generated by GPT to 256 dimensions, which is comparable to the setting in the previous study~\cite{wang2020you}.

\subsection{Threat detection}

\begin{table*}[!th]
  \caption{Attack scenarios involved in \texttt{ProvSec}~\cite{shrestha2023provsec}}\label{tab:vuls}
  \begin{tabular}{clll}
    \toprule
    ID & Software & Vulnerability & Description\\
    \midrule
    \texttt{\#01} & Consul	& N/A	           & Consul service APIs misconfiguration, RCE and reverse shell\\
    \texttt{\#02} & Nginx & CVE-2017-7529	       & Web server remote integer overflow vulnerability\\
    \texttt{\#03} & Ghostscript	& CVE-2018-16509     & Python remote shell command execution via ghostscript\\
    \texttt{\#04} & PHP	    & CVE-2018-19518	    & PHP IMAP remote command execution  (RCE) vulnerability\\
    \texttt{\#05} & Docker	& CVE-2019-5736	     & Escape from a Docker container: vulnerability on Docker\\
    \texttt{\#06} & Tomcat	& CVE-2020-1938	  & Apache Tomcat arbitrary file read / include vulnerability\\
    \texttt{\#07} & Redis	& CVE-2022-0543	   & Redis Lua sandbox escape and remote code execution\\
    \texttt{\#08} & Django	& CVE-2021-35042	& Django allows QuerySet.order\_by SQL injection vulnerability \\
    \texttt{\#09} & Apache	& CVE-2021-42013	& Path traversal and file disclosure vulnerability in HTTP server \\
    \texttt{\#10} & Apache  & CVE-2021-41773	   & Apache web server path traversal and file disclosure vulnerability \\
    \texttt{\#11} & Java Log4j & CVE-2021-44228	     &Log4j vulnerability allows an affected system to be controlled remotely\\
    \bottomrule
  \end{tabular}
\end{table*}

It is necessary to emphasize that the primary focus of this work is to explore the potential of LLMs in improving provenance analysis. We are particularly interested in how the extensive knowledge of LLMs can be transferred to the field of APT detection and play an active role. Therefore, we do not intend to develop sophisticated and complex threat detection systems based on architectures like graph neural networks or Transformers. Moreover, we aim to fully leverage the assistance of LLMs to reduce human effort, such as feature engineering. Consequently, we consider two categories of classic models as our threat detectors: supervised learning methods and semi-supervised anomaly detection methods. 
For the supervised learning, two representative algorithms are included. The first one is multilayer perceptron (\texttt{MLP}), which is a neural network model. The second detector is gradient-boosted decision trees (\texttt{GBDT}), which is a representative ensemble learning model. For the semi-supervised learning, we train an anomaly detector using \texttt{XGBOD}~\cite{zhao2018xgbod} algorithm. This is more aligned with real-world application scenarios, where fully labeled data is hard to obtain, and artificial intelligence applications are built based on a combination of labeled and unlabeled data.

\section{Evaluation}\label{sec:eval}

Based on the methodology introduced in Section~\ref{sec:method}, we conduct a set of experimental studies to evaluate the performance of our proposed technique. We are particularly interested in exploring the interpretability of descriptive tests from GPT-4o regarding system events, as well as the ability of OpenAI's embedding model to capture the semantics of system events. 
and their effectiveness in practical threat detection. We would also like to further investigate whether OpenAI's extensive knowledge can be practically helpful for threat detection. Specifically, our evaluation is intended to seek a deeper understanding of the following research questions (\textbf{RQs}).
\begin{itemize}
    \item \textbf{RQ1:} Do the embeddings generated by LLMs provide a significant advantage in anomaly detection compared to previous methods? (\textbf{\S4.3} and \textbf{\S4.4})
    
    \item \textbf{RQ2:} Can the extensive knowledge of LLMs regarding system events be effectively transferred to the application of APT detection? (\textbf{\S4.5})
    
    \item \textbf{RQ3:} How is the generalization capability of our proposed APT detection technique when confronted with a previously unseen attack? (\textbf{\S4.6})
    
\end{itemize}


\subsection{Dataset}

We conduct empirical studies based on a publicly available dataset \texttt{ProvSec}~\cite{shrestha2023provsec}. This dataset collects provenance analysis data from eleven attack scenarios in the real world. Information about these eleven attacks is summarized in Table~\ref{tab:vuls}. When an attack is launched, the corresponding events are recorded and preliminarily regarded as adversarial. Otherwise, if the attack has not been launched, the events extracted from the corresponding provenance data are regarded as benign. A self-evident intuition is that adversary attacks are usually hidden among massive normal events. Therefore, it is unsurprising that there exist some events that may be on both sides. The intersection between the benign events and the preliminary adversary events is not sufficiently iconic to indicate an attack behavior. To this end, those overlapped patterns from the original adversary events should be removed. Please note that, as described in Section 3, we extracted slightly different fields to depict different types of events. As a result, a total of 4,507 unique adversary events remained. Moreover, in reality, the distribution of adversary events and normal events is imbalanced, with the majority of events being benign. To keep rough balance, we randomly extract 5,000 different benign events. In the dataset, 80\% of instances are used for training and the remaining 20\% for testing, denoted as $\mathcal{D}_{Train}$ and $\mathcal{D}_{Test}$, respectively.


\subsection{Environment Settings}

When interacting with GPT-4o via the API provided by OpenAI, we select
``\texttt{gpt-4o}''
as the LLM. Furthermore, we choose the \texttt{text-embedding-3-small} as our text embedding approach, which is the newest embedding model released by OpenAI in January 2024. 
Compared to its more powerful counterpart \texttt{text-embedding-3-large}, \texttt{text-embedding-3-small} is smaller but still highly efficient. More specifically, we chose 1,536 as the embedding dimension for the output of the \texttt{text-embedding-3-small} model. According to the evaluation results released by OpenAI~\cite{emb_openai}, its performance is comparable to that of \texttt{text-embedding-3-large}.

We implement the prototype system in Python 3. Our system is carried out in the Ubuntu 22.04.1 LTS environment running on a computer equipped with an Intel\textsuperscript{\textregistered} 
Core\textsuperscript{TM} i9 CPU, 64GB RAM. This computer also has a CUDA-based parallel computing platform with a NVIDIA\textsuperscript{\textregistered} GeForce RTX\textsuperscript{TM} 3080 GPU. 

\subsection{Analysis of Embedding Quality}

\subsubsection{Embedding Visualization}

\begin{figure*}[!t]
\centering
\subfloat[Embeddings are generated using Doc2Vec] {\includegraphics[scale=0.54]{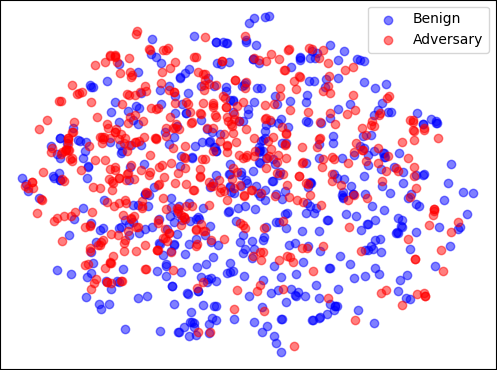}}
\quad\quad\quad
\subfloat[Embeddings are generated using GPT of OpenAI]{\includegraphics[scale=0.54]{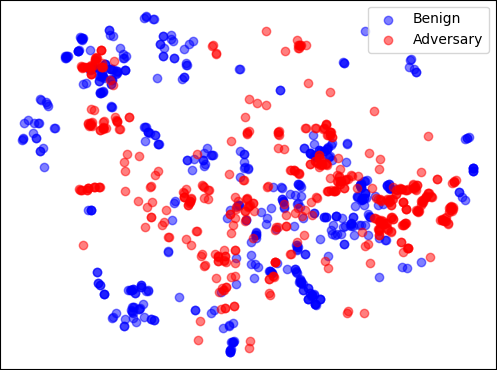}}
\caption{Visualization based on t-SNE of event embeddings. The blue points and red points represent benign events and adversary events respectively.}\label{fig:emb_comp}
\end{figure*}

In this section, we first compare \texttt{Doc2Vec}~\cite{le2014distributed} with GPT-4o in terms of the interpretability of event embeddings. The former method was used in a prior research \texttt{ProvDetector}~\cite{wang2020you}. 
To keep balance, for benign and adversary events, we randomly selected 500 samples from each. After that, we project the event embeddings to a two-dimensional space using t-SNE~\cite{van2008visualizing}. The result is shown in Figure~\ref{fig:emb_comp}, where blue points represent benign samples while red ones represent adversarial samples. 

It can be seen from Figure~\ref{fig:emb_comp}(a) that the majority of malicious events are mixed with benign ones. This is consistent with the result presented in the existing paper~\cite{wang2020you}. To address this challenge, previous researchers design a rareness-based selection algorithm to first extract the potentially malicious part from the provenance graph of a process. However, this method is based on heuristics, which consequently relies on human expertise and typically leads to a risk of bias. 

In contrast, our method leverages the extensive knowledge of GPT-4o to enrich the semantics of event descriptions. Therefore, we can observe from Figure~\ref{fig:emb_comp}(b) that small cliques formed by similar events, although benign and adversarial samples are not trivially linear separable. This is because we compress the high-dimensional space into a 2-dimensional plane, which leads to information loss. The separating hyper-planes between groups
that could originally be distinguished are no longer apparent. However, it also implies that the existing event
embeddings can be used as a feasible starting point. Based on this, we can take advantage of the subsequent
machine learning models to further amplify the distance between benign and adversary events in a feature
space. At the same time, the distance between two benign events, and the distance between two adversarial
events can both be reduced. The experiment results presented in Section 4.4 and Section 4.5 validate our hypothesis.

\subsubsection{Explicability of Embedding Model}

\begin{figure}[!t]
\centerline{\includegraphics[width=0.46\textwidth]{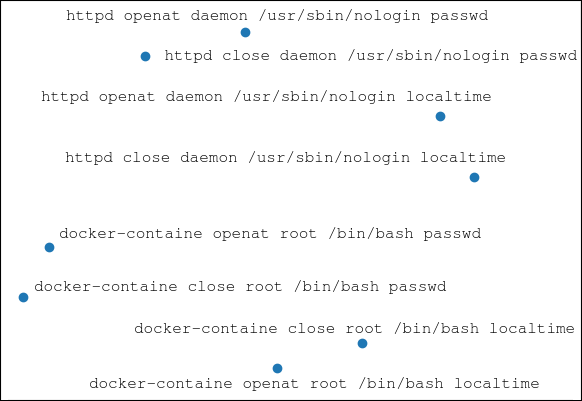}}
\caption{The relative positions of eight system events in the embedding space are visualized using the MDS technique.}
\label{fig:mds}
\end{figure}

Intuitively, if the numerical representations for events provided by the embedding model are semantically meaningful, the degree of similarity between events can be roughly reflected in the embedding space by the proximity or distance. 
To delve deeper into the interpretability of GPT-generated embeddings, we showcase two pairs of system events as case studies. The first pair of system events are as following:

``\texttt{touch execve root /bin/bash <NA>}'' 

``\texttt{sh execve root /bin/bash <NA> touch}''

\noindent Obviously, the processes involved in the two events are completely different. In the latter event, \texttt{sh} is a shell command line interpreter that executes a command specified by the event argument, which in this case is \texttt{touch}. According to our experience, these two events have the identical functionality. The cosine distance between their numerical embeddings is approximately 0.10884, indicating that the inference of the LLM is consistent with our prior knowledge. We also can observe the similar result when inspecting the following events pair:

``\texttt{useradd execve root /bin/bash <NA>}'' 

``\texttt{sh execve root /bin/bash <NA> useradd}''

\noindent The cosine distance between their numerical embeddings is approximately 0.10778. 

\vspace{3pt}
Furthermore, we examine the quality
of GPT-based system event embeddings in terms of their analogical reasoning performance. More concretely, we plot the relative positions of eight events according to their cosine distance in a lower-dimensional space using the Multidimensional Scaling (MDS) technique. The result is shown in Figure 4, from which it is easy to observe that the distance between the following two events

\begin{figure*}[!th]
\centering
\subfloat[MLP-based detector]{\includegraphics[scale=0.38]{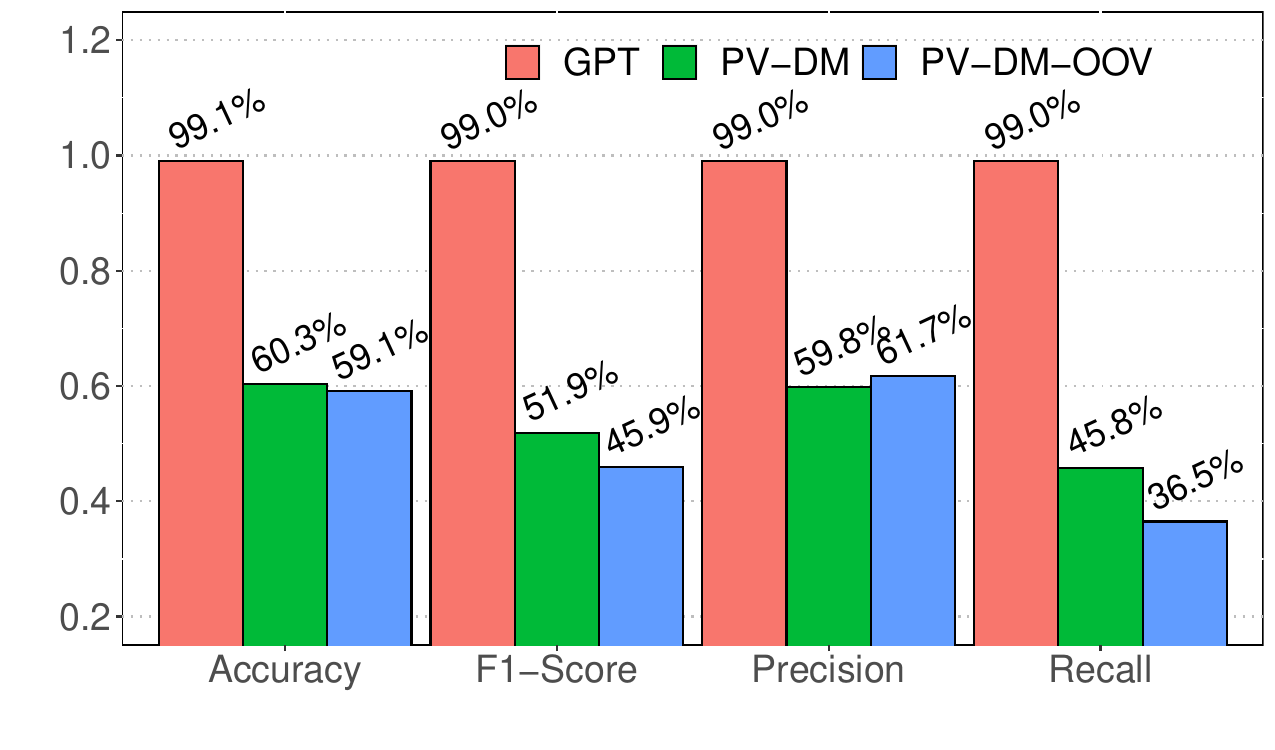}}\ \ 
\subfloat[GBDT-based detector] {\includegraphics[scale=0.38]{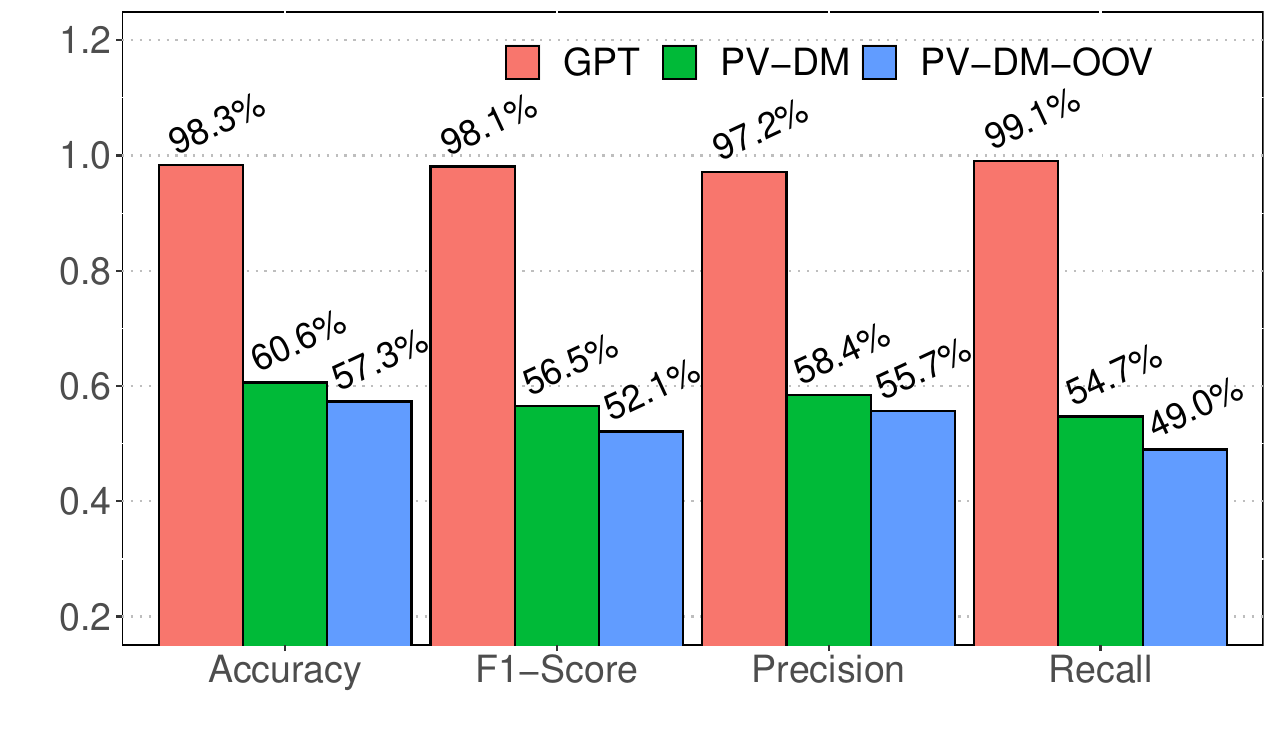}}
\caption{Comparison of different embedding methods for the final threat detection performance.}\label{fig:comp}
\end{figure*}

``\texttt{httpd openat daemon /usr/sbin/nologin passwd}'' 

``\texttt{httpd close daemon /usr/sbin/nologin passwd}'' 

\noindent is close to the distance between

``\texttt{httpd openat daemon /usr/sbin/nologin localtime}'' 

``\texttt{httpd close daemon /usr/sbin/nologin localtime}''

\vspace{3pt}
\noindent Similarly, the distance between the following two events

``\texttt{docker-containe openat root /bin/bash localtime}'' 

``\texttt{docker-containe close root /bin/bash localtime}'' 

\noindent is close to the distance between 

``\texttt{httpd openat daemon /usr/sbin/nologin localtime}'' 

``\texttt{httpd close daemon /usr/sbin/nologin localtime}''

\noindent This parallelogram relationship in geometry is consistent with our prior knowledge. We limit the presented examples to eight due to space limitation. In our manual investigation, however, we find many such semantic analogies that are automatically learned. Therefore, it can be seen that the embeddings demonstrated good explicability.



\subsection{Comparison}

In this section, we compare the representation learning adopted in our approach with the embedding method used in the previous paper~\cite{wang2020you}, and examine their impact on the final threat detection. More concretely, the \texttt{PV-DM} model of \texttt{Doc2Vec} was employed in~\cite{wang2020you}. When using \texttt{PV-DM} to generate embeddings for system events, we consider two different scenarios. The first scenario assumes that we have exhaustively collected as many words as possible for the vocabulary, so there is no out-of-vocabulary (OOV) issue. Surely, this is difficult to achieve in practice. Therefore, the second, more realistic scenario is that we only learn an embedding model from $\mathcal{D}_{Train}$. As such, during the testing phase, there may be words that have never been encountered before. 

It should be noted that, to ensure fairness, we use Kernel PCA to reduce the dimensionality of the feature vectors generated by GPT, from 1,536 to 256. Similarly, the feature vectors generated by \texttt{Doc2Vec} are also kept at the same dimensionality, i.e., 256 dimensions. As mentioned earlier, for the threat detector, we consider two different supervised learning methods here: \texttt{MLP} and \texttt{GBDT}. For MLP, we conduct training for 50 rounds and choose the model that achieves the highest accuracy score. For GBDT, we use grid search to find the parameters that yield the best accuracy score. The experimental results are shown in Figure~\ref{fig:comp}. 

As shown in Figure~\ref{fig:comp}, regardless of which machine learning model was used as the detector, our proposed method achieves significantly better performance compared to the baselines. When using \texttt{MLP} and \texttt{GBDT} as detectors, the accuracy reaches 99.1\% and 98.3\%, respectively. Moreover, when the OOV problem is present, the performance of detectors trained on feature vectors generated by \texttt{Doc2Vec} further deteriorates. It is worth noting that the size of our testing set is relatively small, containing only 1,902 samples. In real-world scenarios, due to the diversity of unstructured data such as file names or executable paths, the OOV problem is likely to be even more severe. The advantages of large language models would become even more apparent.

\begin{table}[!t]
\caption{Threat Detection Performance}
\begin{tabular}{@{}ccccc@{}}
\toprule
Model     & Accuracy & Precision & Recall & F$_1$-Score \\ \midrule
\texttt{MLP}  & 99.1\%   & 99.0\%  &  99.0\% & 99.0\%   \\
\texttt{GBDT} & 98.3\%   & 97.2\%    & 99.1\% & 98.1\%     \\ 
\texttt{XGBOD} & 96.1\%   & 96.9\%    &  94.7\% & 95.8\%  \\ \bottomrule
\end{tabular}
\label{tab_detect}
\end{table}

\subsection{Detection Performance}

In the previous section, we have witnessed the classification performance of two supervised models, i.e., \texttt{MLP} and \texttt{GBDT}. In the supervised learning setting, our method effectively distinguishes adversary events from normal events. Now, we consider a more challenging but also more practical scenario—verifying whether our proposed method can still be effective under a semi-supervised learning setting. In practice, obtaining a fully labeled dataset is inherently challenging and usually expensive. However, semi-supervised learning uses a combination of labeled and unlabeled data to train artificial intelligence (AI) models. Specifically, semi-supervised outlier detection can be performed even when the training data consists only of observations describing normal behavior. After all, in reality, collecting normal system events is often much easier than acquiring malicious system events. We can then predict whether unknown events are caused by an attack by evaluating their deviation from known events. 

In detail, we adopt \texttt{XGBOD}~\cite{zhao2018xgbod} to build a threat detection model, which is a semi-supervised outlier detection algorithm. Specifically, our anomaly detection was implemented based on \texttt{PyOD}~\cite{zhao2019pyod}, an open-source Python toolbox for performing scalable outlier identification. We compared the performance of this model with the two previous supervised learning methods, as shown in Table~\ref{tab_detect}. Not surprisingly, the performance of semi-supervised learning is slightly worse than that of supervised learning. However, the \texttt{XGBOD}-based threat detection model still achieved good detection results, with a precision of 96.9\%. Thus, it can be concluded that the extensive knowledge of LLMs regarding system events can be effectively transferred to the application of APT detection.

\subsection{Case Study: Detecting Unseen Attacks}

We perform a case study in wild to inspect whether the proposed technique can be generalized over previously unseen attacks and make accurate predictions. In detail, we consider the unseen attack through exploiting CVE-2021-44228. This \texttt{Log4j} vulnerability was discovered in the Log4j logging library, which has severe and widespread impacts. Malicious actors can use the \texttt{Log4j} flaw to run almost any code they want on vulnerable systems. 

Following the previous setup, our training set contains 5,000 benign events, and after excluding the \texttt{Log4j} attack, the training set includes 3,681 adversary events. After that, we randomly extract 500 adversary events from \texttt{Log4j} attack as the testing set. We plot a ROC (receiver operating characteristic) curve to graphically illustrate the performance of this APT detector. We also include 500 benign events in this new testing set because the ROC curve is the plot of the true positive rate (TPR) against the false positive rate (FPR) at varying threshold setting. 

\begin{figure}[!t]
\centerline{\includegraphics[width=0.32\textwidth]{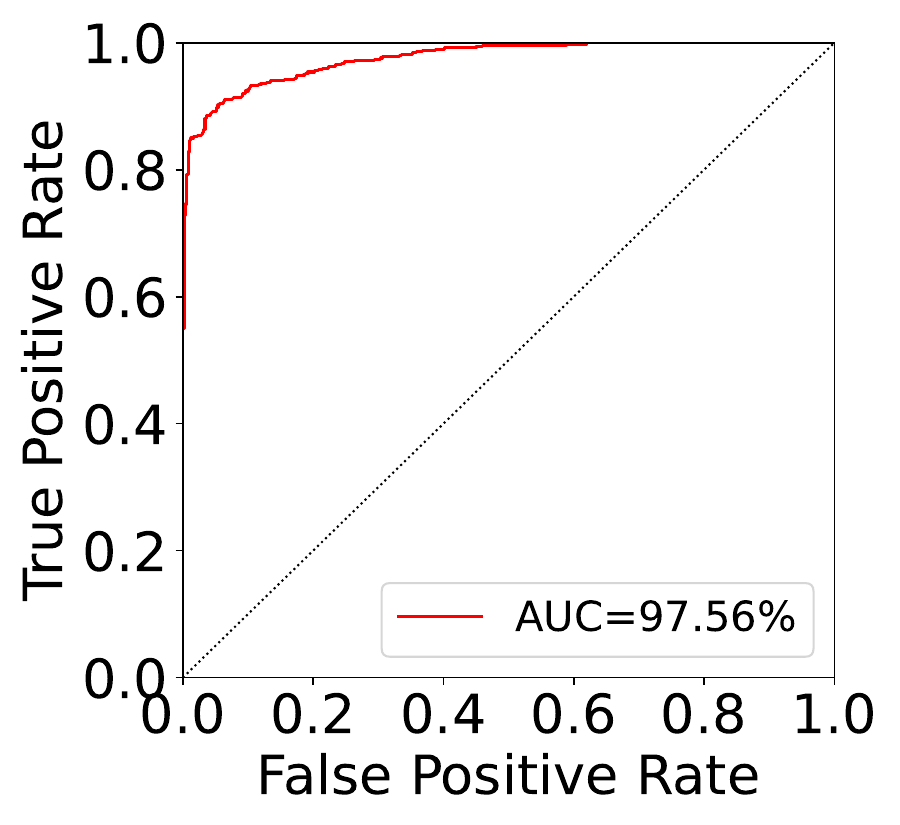}}
\caption{The ROC Curve.}
\label{fig:roc}
\end{figure}

Still, we adopt the semi-supervised algorithm \texttt{XGBOD}~\cite{zhao2018xgbod} to detect adversary events launched by the \texttt{Log4j} attack. The ROC obtained on the testing set is shown in Figure~\ref{fig:roc}, with an AUC value of 97.56\%. This indicates that the proposed technique demonstrates strong adaptability especially when faced with attacks that were never encountered before.

\section{Related Work}\label{sec:related}

In this section, we briefly review the existing literature related to our work. Considering the large volume of research in this
field, the following review is not intended to be exhaustive.

\subsection{Provenance Analysis}

Provenance data capture the relationships and interactions between system entities, providing a comprehensive view of information flow and causal relationships within a system. Therefore, provenance analysis plays a significant role in both real-time threat detection and post-incident forensic investigation. Dependence tracking analysis~\cite{king2003backtracking} has been used to analyze a large volume of data effectively. Provenance tracking has been done in different data granularity. \texttt{BEEP}~\cite{lee2013beep} and \texttt{Protracer}~\cite{ma2016protracer} use units that are execution partitions of application code which is common in event-handling loops. \texttt{MPI}~\cite{ma2017MPI} uses user input on data structures to define execution partitions. \texttt{PrioTracker}~\cite{liu2018priotracker} proposed priority-based causality tracking using rareness score and fanout score as indications of unusualness. Bates et al.~\cite{whole} proposed Linux Provenance Modules (LPM), a kernel-based framework and data loss prevention system for sensitive data. \texttt{PalanTir}~\cite{palantir} uses a processor tracing (PT) hardware technique to enable finer-grained instruction level tracking. \texttt{Kairos}~\cite{cheng2024kairos} proposed a graph neural network-based encoder and decoder to learn the temporal evolution of the provenance graph’s structural changes. \texttt{ProvDetector}~\cite{wang2020you} is the closest related method to the one proposed in this paper. For a causal path consisting of multiple events, \texttt{ProvDetector} uses the \texttt{PV-DM} model of \texttt{Doc2Vec}~\cite{le2014distributed} to learn an embedding. The shortcomings of this approach are evident: first, the semantics it can capture are shallow. However, our method leverages the extensive knowledge of LLMs to enrich the semantics of event descriptions. Secondly, \texttt{ProvDetector} cannot well address the OOV (out-of-vocabulary) challenge. Our approach, on the other hand, does not face this issue.

\subsection{Threat Detection}

Using provenance in threat detection and investigation has been explored by a large body of work~\cite{li2021threat}. Multiple attack detection approaches have been proposed based on anomaly scores. For example, the threat detection system \texttt{PIDAS}~\cite{xie2016unifying} computes the anomaly score of a certain length of a path in a provenance graph and compares it with a predefined threshold. A subsequent work, \texttt{Pagoda}~\cite{xie2018pagoda}, considers the anomaly score of both a single provenance path and the whole graph. Furthermore, \texttt{NoDoze}~\cite{hassan2019nodoze} uses a network diffusion algorithm that propagates anomaly scores
across dependency graphs to calculate anomaly scores. 

Tag propagation is another widely used strategy. \texttt{SLEUTH}~\cite{hossain2017sleuth} introduced two types of tags, namely, trustworthiness tags and confidentiality tags, in the attack
detection system. In a nutshell, an alarm is triggered when a node with low trustworthiness accesses a node with high confidentiality. 
The subsequent work~\cite{hossain2020combating} improved the original tag-based system and thus reduced false alarms significantly in the detection of APT-style attacks. However, tag propagation-based approaches suffer from the ``dependency explosion'' problem. To address this issue, \texttt{Holmes}~\cite{milajerdi2019holmes} prioritizes broad detection using relatively simple signatures to catch a wide range of malicious activity. It then uses alert filtering to reduce false alarms. But \texttt{Holmes} involves many empirical parameters, which leads to unstable detection results.

\subsection{LLMs for Cybersecurity}

Existing work largely employs the GPT family of LLMs to address different security challenges. For instance, previous works~\cite{qu2024context,zhang2024prompt} investigated the capabilities of ChatGPT in detecting vulnerabilities and resolving bugs. In particular,  Qu et al.~\cite{qu2024context} found the debug performance can be remarkably improved after taking advantage of contextual information. Furthermore, Yan et al.~\cite{yan2023prompt} took advantage of GPT-4 to generate descriptive texts for each API call. Based on this, they developed a BERT-based dynamic malware analysis technique. Lastly, the knowledge of LLMs was also applied to detect DDoS attacks~\cite{guastalla2023application, li2024dollm}. To the best of our knowledge, we are the first to propose a technique that leverages the extensive knowledge of LLMs to assist in provenance analysis for APT detection.

\section{Discussion and Future Work}\label{sec:limit}

\textbf{Generality of Sysdig:}
As a widely used security event tracing system, Sysdig event format is well-known to multiple widely used
 LLMs. For instance, as well as OpenAI's models, open-source LLMs, Meta's LLAMA models, also recognize the sysdig format.
The advantage of this event format to LLMs is mainly the clarity of the system event details broken down into clear key-value pairs. Alternatively, for LLMs without the knowledge of Sysdig, the same details of system call event information can be formulated as a prompt.

\textbf{Usage of OpenAI LLM Models:}
At the time of our experiment, OpenAI GPT-4o has the best performance as an industry-leading product. Also, the text embedding models of OpenAI are competitive. Therefore, we chose OpenAI's models for our evaluation. 

However, we note the proposed method is agnostic to LLM models and is also applicable to other LLM models as well. Other open-source or commercial LLM models such as Meta's LLAMA, Microsoft Copilot, and Google Gemini can be utilized for the same purpose.

\textbf{Generality of JSON input:}
We chose the JSON as the input format for LLMs because this is a common format widely utilized by LLMs as an input and output along with other formats such as Markdown.
Since JSON is a structure that clearly delivers multiple key-value pairs, we used it for the brevity of the input prompt. Prior state-of-the-art approaches used the sentence format formulating the system call event as a subject, a verb, and an object format.
This is just an alternative format that is equally effective for LLMs if we explain the positional role of each token such as the first term is a subject, the second term is a verb, etc.
However, this sentence format is likely to make the prompt longer causing a higher cost for the paid LLM models.

\textbf{Future Work:} Our work mainly utilizes the state-of-the-art LLMs of OpenAI for our evaluation because our manual evaluation of the explanation quality is satisfactory.
As our future work, we plan to evaluate other LLMs performance for security tasks. Depending on the number of parameters, the training data size, and the inclusion of security context for training, each LLM model's capability could vary. We look forward to investigating the security expertness of different LLM models.

\section{Conclusion}\label{sec:conclude}

In this paper, we explored a novel APT detection method that offers a significantly improved detection performance by utilizing LLMs to augment the semantics of provenance analysis events.

We found the state-of-the-art LLM offers multiple enhancements over provenance event details. Specifically, our analysis summarizes LLM can offer knowledge on system calls, knowledge on software identity, high-level knowledge on application execution context, and comments on possible suspiciousness beyond a brief description of a system call. Such new details empower the semantic details thus improving the performance of NLP-based detection methods.

Our experiment shows that text embedding on the augmented LLM description significantly improves the performance 
over 
the representative method adopted by a previous work in supervised learning-based detectors. 
Not only that, even in the semi-supervised learning scenario, our anomaly detection can still achieve a precision as high as 96.9\%. This result demonstrates the usefulness of the LLMs for APT detection with regard to the richness of semantic information.

\section*{Acknowledgments}

Sandia National Laboratories is a multimission laboratory managed and operated by National Technology and Engineering Solutions of Sandia, LLC., a wholly owned subsidiary of Honeywell International, Inc., for the U.S. Department of Energy’s National Nuclear Security Administration under contract DE-NA0003525. This article describes objective technical results and analysis. Any subjective views or opinions that might be expressed in the article do not necessarily represent the views of the U.S. Department of Energy or the United States Government. This work was supported through contract CR-100043-23-51577 with the U.S. Department of Energy.

\bibliographystyle{ACM-Reference-Format}
\bibliography{refs}

\end{document}